\documentclass[prl,twocolumn,showpacs,byrevtex]{revtex4}
\usepackage{pifont,graphicx}

\begin{document}

\title{Do shifting Bragg peaks of cuprate stripes reveal fractionally
charged kinks ?}

\author{Marco Bosch}
\email[To whom correspondence should be addressed. Electronic
address:\space]{mbosch@lorentz.leidenuniv.nl}
\author{Wim van Saarloos}
\author{Jan Zaanen}
\thanks{
J. Zaanen acknowledges the hospitality of the Institute of Theoretical 
Physics at the University of California, Santa Barbara where part of
this work was done. This research was supported in part by  the
National Science Foundation under Grant No. PHY94-07194. }
\affiliation{Instituut-Lorentz, Universiteit Leiden, P.O. Box 9506, NL-2300 RA Leiden, The Netherlands}
\date{\today}


\begin{abstract}

The stripe phases found in correlated oxides can be viewed as
an ordering of the solitons associated with doping the Mott-insulating
state. Inspired by the recent observation that the
stripes tilt away from the main axis of the crystal lattice in the
regime $x \approx 1/8$, we propose that a new type of stripe phase is
realized in the large doping regime. This new phase should be viewed
as a doped version of the microscopically insulating $x \lesssim 1/8$ stripes. 
The topological excitations 
associated with the extra doping are fractionally charged
kinks along the stripes whose motions make the stripe fluctuate. 
We argue that the directional degree of freedom of the kinks might
order, causing the stripe phase to tilt. Quantitative predictions
follow for the doping dependence of the tilt angle, which in turn
can be used to determine the
 fundamental charge quantum associated 
with the stripe phase.

\end{abstract}     

\maketitle

In the last few years, overwhelming experimental evidence has
accumulated that the holes in the cuprate materials, such as the high
$T_c$ superconductors, form various types of striped phases 
\cite{tranquadaover,sciencerep,scienceper}. The
building blocks of such phases consist of domain walls which can
be viewed as string-like
configurations of holes that separate hole-free antiferromagnetic
domains. Important information about the microscopic origin of stripes
is contained in the doping dependence of the inverse of the stripe
spacing, shown in Fig.\ 1A. This spacing can be obtained
directly from the the distance  $\delta$  between the measured
Bragg-peaks (Fig.\ 1B).  It was experimentally found that
$\delta$ is to a good approximation directly proportional to doping
$x$ for a variety of stripe systems like the nickelates
\cite{Nickelates} and manganites \cite{Manganites}.  It is also found
in cuprate stripe phases, but only in the doping regime where $x
\lesssim 1/8$, while for $x \gtrsim 1/8$, $\delta$ becomes roughly $x$
independent --- see Fig.\ 1A \cite{Yamadaplot,Uchida}.  In the regime $x<0.05$, the stripes become
diagonal \cite{Wakimoto}. Quite recently it was
found \cite{Matsushita,Lee,Kimura} that in a number of cuprates the
spin peaks tilt away from the $(1, 0)$ or $(0,1)$ axis in reciprocal
space. This puzzling behavior corresponds with a tilt of the stripe
phase away from the main crystal axis of the perovskite planes in real
space and seems characteristic for the $x \gtrsim 1/8$ regime. 
We want to suggest here that this tilt might reflect the presence of
fractionally charged solitons living on the stripes. This notion leads
to an explicit relation between the $\delta$ versus $x$ dependence and 
the $Y$-shift, which, if confirmed experimentally,
allows one to determine the topological charge of stripes.

The $x \lesssim 1/8$ stripes might be viewed as the condensation
of the topological defects associated with doping the half-filled
Mott-insulator \cite{ZaGu,SteveK,Pryadko} (see Fig.\ 2A).  We here suggest that the 
$x \gtrsim 1/8$ state should be viewed in
turn as the condensation of the topological defects associated with
doping the stripe state itself.  This state is as indicated in Fig.\
2B.  The stripes for $x \gtrsim 1/8$ are locally not different from the $x
\lesssim 1/8$ stripes sketched in Fig.\ 2A, except that once in a while the
stripe `steps sideways': it forms kinks. These kinks carry half the
fundamental charge quantum associated with the stripe-insulator of the
$x \lesssim 1/8$ regime and as they move a stripe always in the same
direction they cause the orientation of the stripe phase to deviate
from the lattice axis, explaining the Y-shift.
Although not based on phase separation between two different types of stripe fillings,
our proposal shares the essential idea, that the crossover around $x=1/8$ and the charge
density and orientation of stripes are related, with a suggestion of 
White and Scalapino \cite{WhiteScal}.

Our central assumption is that the cuprate stripe phase in the $x \lesssim
1/8$ regime is insulating on the {\em microscopic} scale.
Experimentally, the cuprate stripe phases are metallic-like and a
popular viewpoint is that the electronic state on the stripes is
metallic. However, the $\delta = S x$ relation, 
where $S$ is the linear slope in 
the small $x$ limit, argues
strongly against this internal metallicity.  Its meaning, already
suggested by earlier mean field calculations, is simply that every
hole stabilizes a piece of charged stripe with a length which is an
{\em integer multiple} of the lattice constant. This special stability
when the electron charge and the lattice are commensurate, shows that
the internal electronic state of a stripe is insulating on short
length scales. At the same time, stripes which are internally
insulating on microscopic scales are not necessarily inconsistent with
metallic behavior on macroscopic scales. The stripe metal might 
correspond with a quantum-disordered stripe insulator, characterized by a
growth of the quantum fluctuations when length- and time scales are
increasing.  Although we will not address this issue in any further
detail, the remainder should be read as a suggestion for a
critical experimental test of this hypothesis.

\begin{figure}[t]
\includegraphics[angle=-90,width=10cm]{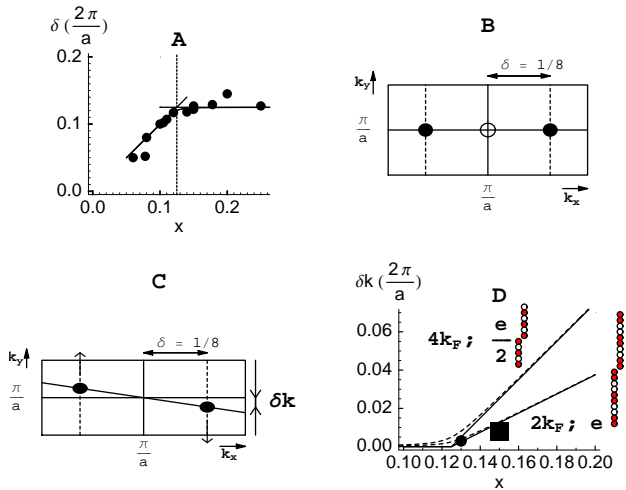}
\caption{\small
({\bf A}) The inverse distance  $\delta$ between the stripes as a function of doping $x$ as measured by Yamada {\em et
al.} \cite{Yamadaplot}. For $x<0.05$, stripes are diagonal. The straight lines are for a
sharp transition at $x=1/8$, the curve for a smooth crossover.
({\bf B})
Sketch of the position of the incommensurate peaks in Fourier-space as measured by neutron-scattering.
The open circle shows where the peaks would have been if the system 
would have been a commensurate antiferromagnet. We only show the peaks resulting from one layer.
({\bf C})
We predict  a shift $\delta k$ of the Fourier-peaks caused by the existence of
kinks in stripes. Because of the resulting tilting of the unit-cell, the peaks would no longer be on the axis,
but move along the dashed lines. The horizontal distance to the $(\frac{\pi}{a},\frac{\pi}{a})$ position stays at
$\delta = \frac{1}{8}$.
({\bf D})
Prediction of the magnitude of
the shift $\delta k$ as a function of $x$ for two possible kinds of stripes: 4$k_F$ stripes with kinks with
fractional charge $\frac{e}{2}$ and $2k_F$ stripes with kinks with charge $e$. The curves
belong to the same smooth crossover as the one shown in Fig.\ 1A. Experimental points:
\ding{108} Matsushita {\em et al.} \cite{Matsushita};
\ding{110} Lee {\em et al.} \cite{Lee}. }
\end{figure}

We follow the literature with regard to the nature of the reference insulator 
in the $x \lesssim 1/8$ regime \cite{WiNa,ZaOl}. 
The slope $S$ of  the $\delta$ versus
$ x$ relationship for small $x$ implies  that one hole stabilizes two charge
stripe unit cells. Accordingly, the dynamics associated with the 
electrons along the stripe can be associated with that of a quarter
filled one-dimensional fermion system \cite{ZXShen}. It is well known that
such a system can become Mott-insulating by breaking translational
symmetry with real space charge periods $2a$ and $4a$, or wavevectors 
$4k_F$ and $2k_F$ respectively, where $k_F
= \pi / (4a)$ ($a$ is the lattice constant). Schematically, the ordering
pattern on the stripe is like  $\cdots - 0 - \updownarrow - 0 - \updownarrow - 
\cdots$ and  $\cdots - 0 - 0 - \updownarrow - \updownarrow - 0 - 0 - \updownarrow - \updownarrow - \cdots$
for the $4k_F$ and $2k_F$ stripes, respectively. Here $0$
denotes the hole and $\updownarrow$ the presence of a spin, which we expect to be disordered due to  quantum fluctuations.
At present it is not known which type of density wave order is realized
on the stripe. All that really matters for the remainder of our discussion
is that the $4k_F$ stripe should be considered as the
condensation of the charge $e$ of one hole \cite{WiNa}, while the
$2k_F$ stripe is associated with the charge quantum of a {\em pair}
of holes, $2e$ \cite{ZaOl}.

Let us now turn to the $x \gtrsim 1/8$ regime. The (near) independence of
$\delta(x)$ as function of doping $x$ implies that a fraction of the
holes $x'=x-\delta(x)/S \approx x - 1/8$ cannot be `absorbed' by the $x \lesssim 1/8$
insulator.  These excess holes should dope the $x \lesssim 1/8$ state and
it is expected that these holes dope the `soft' insulator associated
with the density wave on the charge stripes instead of the magnetic
domains, the remnants of the `hard' insulator of half-filling. Doped
charge density wave states are well understood in the context of
conventional 1D systems \cite{Haldane}.  A key concept was introduced by Schrieffer \cite{Schrieffer} in the
study of polyacetyleen:
the elementary excitations in such a system are not electrons but in
fact parts of an electron.  The reason is that these excitations can
be viewed as electrons bound to topological defects  in the
density wave order parameter.  That the charge of the electron
fractionalizes in doped $4k_F$ density wave systems is easily seen by
considering the strong coupling limit \cite{ZaOsvS}. The undoped state
with one (static) hole added at the central site is indicated in Fig.\ 3A.
After a couple of hops a configuration is reached where the bare hole
has decayed into two propagating excitations carrying half the hole
charge ($e/2$) which are at the same time domain walls (kinks) in the
density wave (Fig.\ 3B). Such an $e/2$ domain wall corresponds with a region of
enhanced charge density, and hence an increased Coulomb energy. In
contrast to a one-dimensional crystal, a stripe has the additional
freedom of moving sidewards, thereby increasing the distance between
the two holes associated with the domain wall \cite{ZaOsvS},
 as sketched in Fig.\ 3C. Thus,
the fractionally charged domain walls also cause the stripe to step
sidewards, modulating the position of the stripes in space, see Fig.\
3D. If all the kinks are in the same direction,
the net result is that the stripe takes on average an orientation
in space which is tilted away from the lattice axis, very much like
slanted phases that are found in lattice string models \cite{eskes}.

\begin{figure}[t]
\includegraphics[angle=-90,width=9cm]{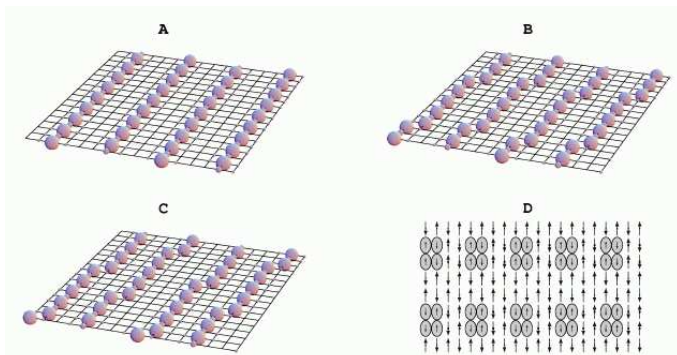}
\caption{\small
({\bf A}) A system of half filled, site-centred 4$k_F$ stripes with doping $x=\frac{1}{8}$ as calculated
in a mean-field Hubbard model. The radius of the spheres denotes the hole-density at a site. The
spin-density is not shown. 
({\bf B}) The same kind of stripe, but now at a doping of $x=0.139$, so $x$ is bigger
than $\frac{1}{8}$. This system has four more holes than the system in (A). The four extra holes are
localised in 8 kinks, so every kink has a fractional charge of $\frac{e}{2}$. ({\bf C}) The same system
as (B), only now the Coulomb repulsion has been reduced by making a Wigner crystal.
 ({\bf D})  A system of 2 half filled, bond-centred 2$k_F$ stripes. The radius of the circles gives the hole-density, the length of the arrows gives the local
magnetisation. This figure is {\em very} similar to a recent DMRG calculation by White and Scalapino
\cite{WhiteScal} on a $t-J$ model. The overall features of the two figures are identical, except near
the boundaries, because of the free boundary conditions used in the DMRG calculations. Notice that every
cluster of four big circles contains in total  approximately two holes, with a singlet orientation for the
spins: real-space Cooper pairs.
}
\end{figure}

Let us now consider a dense {\em system} of such `slanted' stripes. In
the presence of any stripe-stripe interaction this will condense at
zero temperature in a stripe phase where all the stripes are tilted in
the same direction: our explanation for the Y-shift.  It is also
expected that the kinks themselves order in a regular pattern at zero
temperature. The argument is the usual one: the kink gas on a single
stripe becomes at long wavelength a Luttinger liquid showing algebraic
long range order. In the presence of any interaction between the
Luttinger liquids on different stripes this will change into true long
range order at zero temperature \cite{LiqCr}.

The above considerations do not depend on the type of stripe density
wave order, except that the charge $q$ of the kinks is different. How
should the charge of the kinks be counted for $2k_F$ parent stripes?
The rule is that the charge of the soliton is {\em half} the charge
quantum associated with the parent. Since the $2k_F$ stripes can be
considered as condensations of pairs of electrons,
the charge associated with a charged kink in the $2k_F$ stripe becomes
one electron charge. In terms of the strong coupling cartoon of Fig.\ 3,
one should now associate two sites of the real lattice with one site
in the lattice of Fig.\ 3C, so that Fig.\ 3D can be mapped back to the
$2k_F$ case by inserting the doubled lattice and the doubled charges.

These general ideas can be illustrated with explicit calculations.
Since we are considering fully ordered insulating states, mean-field
theory is able to provide meaningful qualitative outcomes. The
reference $2k_F$ and $4k_F$ stripe insulators actually both exist in
the Hartree-Fock solutions of the Hubbard model as low lying, but weakly
metastable states: the true mean-field ground state corresponds with
filled stripes (1 hole per stripe unit cell) \cite{ZaOl,Seibold}.
In Fig.\ 2D we show a typical example
of a straight bond centred $2k_F$ stripe, calculated with the mean-field approximation,
 which strongly resembles the
stripe patterns found in the density matrix renormalization group
calculations by White and Scalapino for the $t-J$ model \cite{WhiteScalDMRG}.

By constraining the distance between the stripes to be fixed,
increasing the hole density and using appropriate boundary conditions,
 doped stripes can also be investigated
in mean-field theory. We find that such doped stripes, if they 
exist as locally stable mean-field solutions, prefer localized
fractionally charged kinks. 
A typical example is shown in Fig.\ 2B where the
density of excess holes is indicated for a $4k_F$ stripe state at
a hole density of $x = 0.139$. In line with the qualitative arguments
given above,
every hole doped into the $4k_F$ $x=1/8$ reference insulator gives rise
to two kinks, each carrying a charge of $e/2$ and moving the stripe
sideways by one lattice constant. Although these `slanted' stripes are
quite stable solutions if the reference insulator is of the $4k_F$
variety, we did not manage to find stable solutions for the $2k_F$
stripes doped with additional holes --- within mean field theory, such 
kinks tend to disintegrate.

\begin{figure}[t]
\centering
\includegraphics[angle=-90,width=9cm]{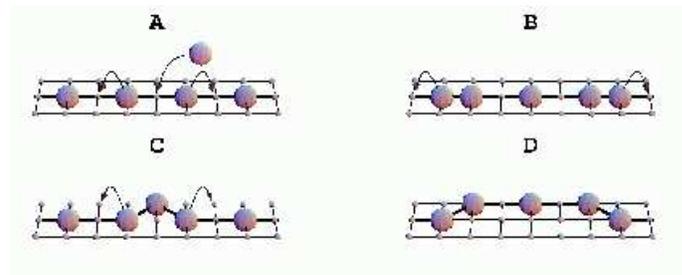}
\caption{\small
Soliton dynamics in a strongly coupled
doped $4k_F$ stripe. The small spheres denote the lattice points with one electron, the large
spheres are holes. The spins are not shown. 
({\bf A}) The reference state: localised stripe with $4k_F$ charge density wave, onto which we add an extra
hole. 
({\bf B}) If the
stripe is rigid, the doped hole separates in a left- and right moving soliton, both carrying half the electron
charge. 
({\bf C}) When the curvature energy becomes less than the charge compressibility energy, the hole can
escape `sideways'. 
({\bf D}) As a result, the stripe now has two kinks that can move independly of each other.
}
\end{figure}

In order to get some feeling of the relative stability of the $4k_F$
slanted phase we compared it with a straight stripe where an additional
hole is simply inserted in the straight density wave ($\cdots -
\updownarrow - 0 - 0 - 0 - \updownarrow \cdots$). On the Hartree-Fock
level, a stripe with two kinks of charge $e/2$ always has
significantly lower energy (of order 0.01 $t$). We also checked that the
sidesteps can be further stabilized by including a nearest-neighbour
Coulomb interaction. The energetics associated with the relative
transversal orientation of the kinks is a more delicate matter which
will strongly depend on the distance between kinks. We believe that
collective fluctuation effects will promote long range slanted order
over zig-zag type stripe patterns, in much the same way as quantum
fluctuations make lattice strings directed \cite{eskes}.

Of course, our calculations merely serve to illustrate the principle
--- neither the question which possibility is realized in nature nor
the question whether the `topological' insulator is the proper
reference state can be settled by these calculations.  However,
simple predictions follow which at least in principle can be checked
by experiment. Most importantly, the charged kinks offer an
explanation for the Y-shift which has already been observed in
experiments. However, the kink notion implies that the superlattice
peaks of the slanted stripe phase are located along the straight lines
crossing the superlattice peaks associated with the $1/8$ phase as
indicated in Fig.\ 1C. The reason is that the average distance between
the stripes does not change when the stripes are making sidesteps,
thereby leaving the distance between the peaks in reciprocal space
along the horizontal axis in Fig.\ 1C unchanged.

Lattice commensuration effects make a transition to 
kinked stripes around $x=1/8$ most likely. However, due to complications
associated with quenched disorder we have no prediction for
whether there should be a sharp transition or a rapid but smooth cross-over. However, the scenario
that there are no free holes, but that instead all holes go onto stripes, leads to an immediate relation between
the $\delta(x)$ dependence and the $Y$-shift. 
Let us define the magnitude of the $Y$-shift as $\delta k$ and the fractional charge (in units of $e$) 
carried by the kink as $q$.
Then $x' =x-\delta (x)/S \approx x-1/8$ is the hole density associated with doping the stripes in the $x \gtrsim 1/8$
regime. It can be shown
that the Y-shift obeys $\delta k = \frac{x'}{2 q} \frac{2 \pi}{a}$ and that $\delta=(x-x')\frac{2 \pi}{a}$
where $a$ is the lattice constant. So we have $\delta+2 q \: \delta k=\frac{2 \pi}{a} x$, independent of
whether there is a crossover or sharp transition, and where it takes place. In Fig.\ 1D we show our prediction
for the Y-shift as function of $x$ for the $2k_F$ and $4k_F$
cases, both for the case of a sharp transition at $\frac{1}{8}$ and for the same smooth crossover as in Fig.\
1A. 
We have also indicated in this figure two experimental results
for the Y-shift, where the $x = 0.15$ point has been reported by Lee
{\em et al.} \cite{Lee} for an oxygen doped sample of $La_2CuO_4$ and the $x=0.13$
point is measured in a $La_{2-x} Sr_x CuO_4$ sample by Matsushita {\em
  et al.} \cite{Matsushita}.  
There is also one experimental indication of a Y-shift at
dopings slightly less than 1/8 by Kimura {\em et al.} \cite{Kimura}, which suggests that there is a smooth
crossover. Clearly an extensive experimental study is called for to
establish whether the Y-shift is due to kinks and
whether our relationship between  $\delta k$ and $\delta (x)$ holds.
The existing data seem to favour a kink charge of $e$ over charge
$e/2$, suggesting that the elementary charge quantum associated with
the stripe phase is $2e$.  Given the proximity of a superconductor
characterised by charge quantum $2e$, such a stripe
charge-quantisation would obviously be quite interesting \cite{footnote2}.

We argued that the charged kinks themselves should order.  It should
therefore in principle be possible to observe the superlattice
reflections of this kink lattice in diffraction experiments. As we
argued, the kinks on single stripes will repel but the kink order from
stripe to stripe is less easy to establish. On the one hand, the kinks
carry charge and under the assumption that the screening length is of
order or larger than the average kink separation the kinks would tend
to maintain a maximum separation, thereby forming a Wigner crystal as
indicated in Fig.\ 2C.  However, one could imagine that elastic
deformation energies are minimised when kinks line up as indicated in
Fig.\ 2B.  The precise realization of the kink superlattice is
therefore a subtle, quantitative matter. Observation of this kink
superlattice may be a formidable experimental challenge.
However, given the conceptual implications implied by such an 
observation it should be given a high priority.

In conclusion, inspired by the tilting of the stripes seen at higher
dopings we have proposed a scenario  of stripe phases characterized
by a proliferation of fractionally charged solitons. Remarkably, by
studying merely the structural characteristics of this phase which is realized
at higher dopings, the elementary charge quantum numbers of the stripes
can be established. This is undoubtedly a crucial piece of information in the
quest for the understanding of high $T_c$ superconductivity \cite{WiNa}.


\vspace*{2cm}



\begin{thebibliography}{27}

\bibitem{tranquadaover}
V. J. Emery, S. A. Kivelson,  J. M. Tranquada, {\it Proc. Natl. Ac. Sci. USA}
 {\bf 96}, 8814 (1999).

\bibitem{sciencerep}
R. F. Service, {\it Science} {\bf 283}, 1106 (1999).

\bibitem{scienceper}
J.~Zaanen, {\it Science} {\bf 286}, 251 (1999).

\bibitem{Nickelates}
J. M.~Tranquada, D. J. Buttrey, V. Sachan, {\it Phys. Rev. B} {\bf 54}, 12318
(1996).

\bibitem{Manganites}
M.~Hervieu {\it et al.}, {\it Eur. Phys. J. B} {\bf 8}, 31 (1999);
S. Mori, C. H. Chen, S-W. Cheong, {\it Phys. Rev. Lett.} {\bf 81}, 3972 (1998).


\bibitem{Yamadaplot}
K.~Yamada {\it et al.}, {\it \prb}   \textbf{57}, 6165 (1998).


\bibitem{Uchida}
The special status of  $x = 1/8$ was recently confirmed by 
magneto-transport measurements: see T.~Noda,
  H.~Eisaki,  S.~Uchida, {\it Science} {\bf 286},  265 (1999).

\bibitem{Wakimoto}
S. Wakimoto {\it et al.}, {\it Phys. Rev. B} {\bf 60}, R769 (1999) and
S. Wakimoto {\it et al.}, cond-mat/9908115.

\bibitem{Matsushita}
 H.~Matsushita {\it et al.},
 {\it J. Phys. and Chem. of Solids} {\bf 60}, 1071 (1999).



\bibitem{Lee}
 Y.~Lee {\it et al.}, {\it Phys. Rev. B} {\bf 60}, 3643 (1999).


\bibitem{Kimura}
 H.~Kimura {\it et al.}, {\it Phys. Rev. B} {\bf 59}, 6517 (1999).



\bibitem{ZaGu}
 J.~Zaanen, O.~Gunnarson,
 {\it \prb} \textbf{40}, 7391 (1989); J. Zaanen, {\it J. Phys. Chem.} {\bf 59}, 1769 (1998). 

\bibitem{SteveK}
S.~A.~Kivelson, V.~J.~Emery, {\it Synth. Met.} {\bf 80}, 151 (1996).

\bibitem{Pryadko}
L.~P.~Pryadko, S.~A.~Kivelson, V.~J.~Emery, Y.~B.~Bazaliy, E.~A.~Demler,
{\it Phys. Rev. B} {\bf 60}, 7541 (1999).


\bibitem{WhiteScal}
S.~R. White, D.~J. Scalapino, {\it Phys. Rev. Lett.} {\bf 81}, 3227 (1998).


\bibitem{WiNa}
C.~Nayak, F.~Wilczek, {\it \prl} \textbf{78}, 2465 (1997); 
C. Nayak, F. Wilczek, \emph {Int. J. Mod. Phys. B} \textbf{10}, 2125 (1996).


\bibitem{ZaOl}
J.~Zaanen, A. M.~Oles, \emph{Ann. Phys. (Leipzig)} \textbf{5}, 224 (1996).

\bibitem{ZXShen}
In recent photoemission experiments evidence is found for an underlying one
  dimensional bandstructure characterized by a Fermi momentum of a
  quarter-filled system: X.~J. Zhou {\it et al.}, {\it Science} {\bf 286}, 268 (1999).

\bibitem{Haldane}
F. D. M. Haldane, {\it Phys. Rev. Lett.} {\bf 45}, 1358 (1980).

\bibitem{Schrieffer}
J. R. Schrieffer, {\it Proceedings of the  International  School of Physics Enrico Fermi LXXXIX}, edited
by F. Bassani (Elsevier, New York, 1985);
S.~Kivelson, J.~Schrieffer, {\it \prb} \textbf{25}, 6447 (1982).


\bibitem{ZaOsvS}
 J.~Zaanen, O.~Y. Osman, W.~van Saarloos,
{\it   \prb} \textbf{58}, R11868 (1998).


\bibitem{eskes}
 H.~Eskes,
  R.~Grimberg,
  W.~van Saarloos,
  J.~Zaanen,
   {\it \prb} \textbf{54}, R724 (1996).

\bibitem{LiqCr}
 We notice that the argument of Kivelson {\em et al.} 
(S. A.~Kivelson, E.~Fradkin, V. J.~Emery,
 {\it Nature} \textbf{393}, 550 (1998)) for the
  dephasing of density correlations on different stripes by transversal stripe
  fluctuations does not apply here. Although valid in the continuum, the kinks
  we consider are tied to the presence of an underlying crystal lattice and it
  is easily seen that the dephasing does not occur.


\bibitem{Seibold}
In Gutzwiller mean-field (qualitatively similar but more accurate than
Hartree-Fock) half-filled stripes become the ground state of the
Hubbard model with added long range Coulomb interaction: see 
 G.~Seibold, C.~Castellani, C.~D.~Castro, M.~Grilli,
 {\it  \prb} \textbf{58}, 13506 (1998).

\bibitem{WhiteScalDMRG}
S. R. White, D. J. Scalapino, cond-mat/9907375.

\bibitem{footnote2}
We also notice that the $4k_F$ density wave is saturated with kinks already at 
$x=1/4$ with $\delta k =1/8 (\pi/a)$. This corresponds with filled stripes oriented along the
diagonals (like the $x=1/4$ stripes in nickelates). 
On the other hand, the $x=1/4$ kink saturated $2k_F$ stripes would show 
$\delta k = 1/16 (\pi/a)$. 
Despite the fact that the peaks are broadening rapidly as function of
increasing $x > 1/8$ we expect that these gross differences should be 
observable.

\end{thebibliography}
\end{document}